\documentclass[10pt]{article}

\usepackage{amsmath}
\usepackage{amssymb}

\usepackage{graphicx}

\usepackage{cite}

\usepackage{color} 

\usepackage[labelfont=bf,labelsep=period,justification=raggedright]{caption}

\makeatletter
\renewcommand{\@biblabel}[1]{\quad#1.}
\makeatother

\date{}

\usepackage{bbm,booktabs}

\begin{document}

\begin{flushleft}
{\Large
\textbf{Fast and Flexible Selection with a Single Switch}
}
\\
Tamara Broderick$^{1,\ast}$, 
David J. C. MacKay$^{2}$
\\
\bf{1} Department of Statistics, University of California, Berkeley, California, USA
\\
\bf{2} Cavendish Laboratory, University of Cambridge, Cambridge, United Kingdom
\\
$\ast$ E-mail: tab@stat.berkeley.edu
\end{flushleft}

\section*{Abstract}

Selection methods that require only a single-switch input, such as a button click or blink, are potentially useful for individuals with motor impairments, mobile technology users, and individuals wishing to transmit information securely. We present a single-switch selection method, ``Nomon,'' that is general and efficient. Existing single-switch selection methods require selectable options to be arranged in ways that limit potential applications. By contrast, traditional operating systems, web browsers, and free-form applications (such as drawing) place options at arbitrary points on the screen.
Nomon, however, has the flexibility to select any point on a screen. Nomon adapts automatically to an individual's clicking ability; it allows a person who clicks precisely to make a selection quickly and allows a person who clicks imprecisely more time to make a selection without error. Nomon reaps gains in information rate by allowing the specification of beliefs (priors) about option selection probabilities and by avoiding tree-based selection schemes in favor of direct (posterior) inference. We have developed both a Nomon-based writing application and a drawing application.
To evaluate Nomon's performance, we compared the writing application with a popular existing method for single-switch writing (row-column scanning).
Novice users wrote 35\%
faster with the Nomon interface than with the scanning interface. An experienced user (author TB, with $>$ 10 hours practice) wrote at speeds of 9.3 words per minute with Nomon, using 1.2 clicks per character and making no errors in the final text.


\section*{Introduction}

In single-switch communication, user input consists of repeated clicks, distinguished only by timing information; these clicks might be generated by pressing a button or blinking. For instance, the range of movement of individuals with severe motor impairments may be limited to a single muscle. Alternatively, a crowded or jostled mobile technology user may be able to click precisely while other actions are difficult or sloppy. A single switch may also be useful when information conveyed, such as a PIN, is sensitive and hand location on a normal keyboard might betray this content. Our method, Nomon (Figures~\ref{fig:ng}, \ref{fig:nk}), expands the application scope of existing methods and facilitates faster writing than the most common single-switch writing interface. 

Existing single-switch communication methods include scanning~\cite{damper:1984, simpson:1999, evreinov:2004, szeto:1993, baljko:2006, venkatagiri:1999, koester:1994, koester:1996, lesher:1998:optimal} and One-Button Dasher~\cite{mackay:2004, mackay:2006, mackay:2007, mead:2009, ward:2000, ward:2002}. (Morse Code does not fall under the strict definition of a single switch interface since it requires either click duration information or multiple switches.) Scanning is the most popular single-switch selection method. In a scanning interface, options such as letters are arranged in a grid (Figure~\ref{fig:grid}). For standard row-column scanning, each row of the grid is highlighted in turn, with the highlight moving to the next row at fixed time intervals, a.k.a.\ {\em scanning delays}. When a click is made, the columns of the selected row are then highlighted in turn, typically iterating at the same fixed time intervals. To select a column, and thereby make a final selection, the user clicks when the highlight is on that column. A variety of customizable commercial scanning software exists for writing and computer navigation~\cite{sensory:2008,words:2000,words:2004,sun:2004} although customization is often not single-switch accessible. The Gnome Onscreen Keyboard~\cite{sun:2004}, by contrast, can generate a grid for new applications ``on the fly.''

While the scanning method can be used to select anything that can be arranged in a grid, One-Button Dasher is limited to writing with alphabetic character sets. Dasher works by arranging all possible character strings in alphabetic order and having the user zoom in on the desired string. More likely strings, according to the language model, are given relatively more space and are thus easier to select.

Scanning and One-Button Dasher require options to be arranged in a particular configuration. By contrast, traditional operating systems, web browsers, and free-form applications such as drawing place options at arbitrary points on the screen.
Scanning, the most popular single-switch communication method, is limited in further ways by its grid structure. For instance, the grid options may theoretically be reordered after any selection to allow the most likely options to be selected the most quickly. However, in practice this reordering requires that users either learn many grid arrangements or search the grid for their desired option upon each reordering.

Even scanning a grid that maintains a fixed layout at all times has drawbacks. Previous studies suggest that, at least among children, scanning a fixed grid demands a higher cognitive load than direct selection~\cite{horn:1996,mizuko:1994,ratcliff:1994,wagner:2006}---though an earlier study found no difference~\cite{mizuko:1991}. One implicated factor is the need for a user to divide her attention between the scanning highlight and the desired option~\cite{horn:1996,wagner:2006}. Another issue in scanning is the possibility of distraction, and loss of the target from working memory, while highlighting progresses~\cite{petersen:2000,wagner:2006}.

Therefore, we seek a single-switch selection method that is not limited to certain forms of option placement. We want our method to work for any number of options; to be able to effectively reorder the set of selections without imposing additional cognitive load; and to allow the user to attend only to the desired target.

Below, we begin by describing such a method, which we call ``Nomon.''
We also describe how our method can adapt to individuals' clicking abilities and how it can incorporate prior beliefs about option selection frequency. In order to evaluate our method's performance, we note that much single-switch research has focused on optimizing writing speed~\cite{damper:1984, simpson:1999, evreinov:2004, szeto:1993, baljko:2006, venkatagiri:1999} and the number of clicks per output symbol~\cite{koester:1994, koester:1996, lesher:1998:optimal} in scanning interfaces. In light of these studies, we developed a writing application, the Nomon Keyboard (Figure~\ref{fig:nk}), using our method and compared its performance with a popular commercial scanning interface, The Grid 2~\cite{sensory:2008} (Figure~\ref{fig:grid}). We examined the study participants' writing speeds, error rates, and number of clicks made per character as well as the subjective ratings of their experiences.

The full technical report describing Nomon is available online at \linebreak {\tt http://www.inference.phy.cam.ac.uk/nomon/files/nomon\_tech\_report.pdf}. The Nomon Keyboard, as well as a drawing application (Nomon Draw) and instructions for the use of both applications, is available for download at {\tt http://www.inference.phy.cam.ac.uk/nomon/} under the GNU General Public License 3.0.


\subsection*{A New Method}

\begin{figure}[!t]
\begin{center}
\setlength{\fboxsep}{0pt}
\fbox{\includegraphics[scale=1]{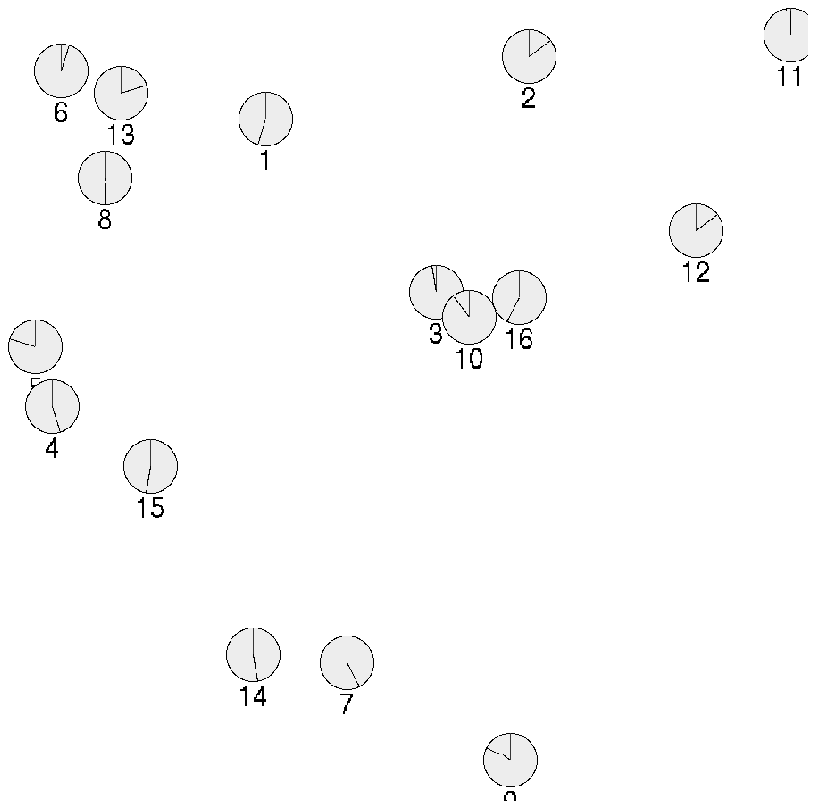}}
\caption{{\bf An example Nomon application for selecting between 16 points on screen (screenshot).} The horizontal and vertical positions of the option points were chosen uniformly at random in the box shown to illustrate the flexibility of the method.}\label{fig:ng}
\end{center}
\end{figure}

\begin{figure}[!t]
\begin{center}
\includegraphics[scale=1]{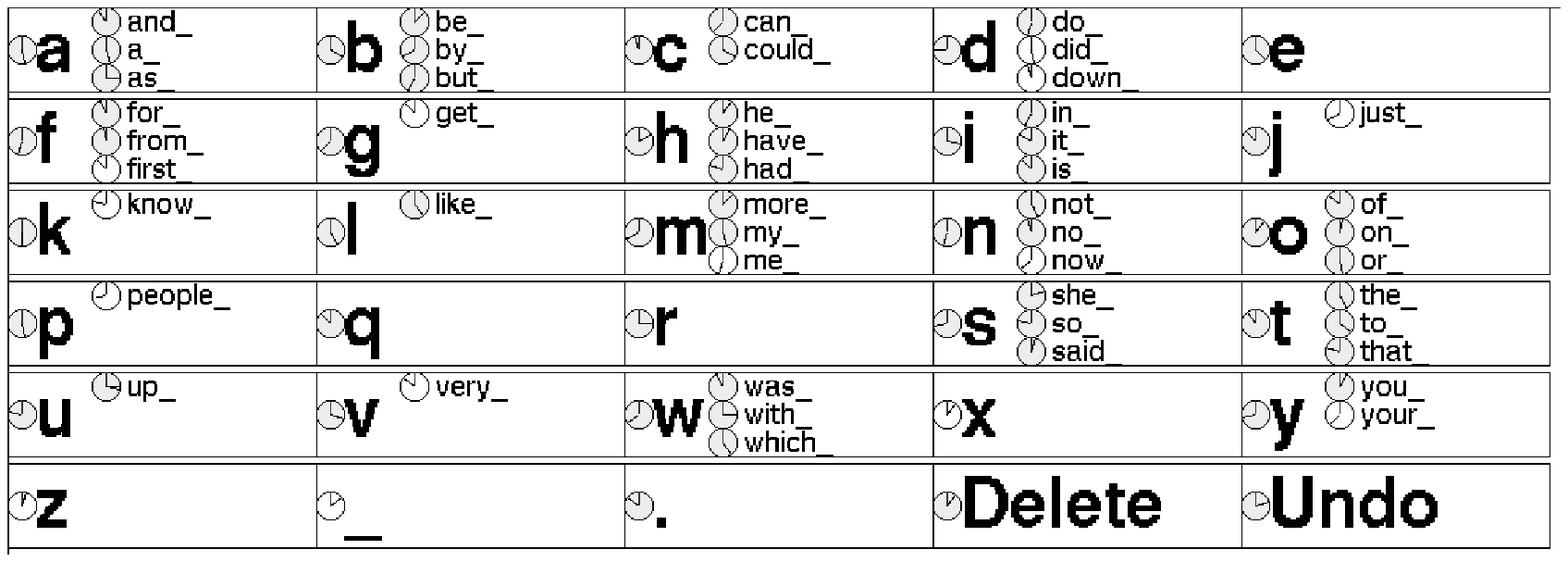}
\caption{{\bf The Nomon Keyboard, a writing application (screenshot).} Words that are prefixed by the concatenation of the current context and the letter X appear next to the letter X. Underscore represents a space. Options for period, a character-deletion function, and an undo function are also available.}\label{fig:nk}
\end{center}
\end{figure}

Nomon, a new single-switch communication method, does not limit the user to selecting options that can be arranged in a grid or alphabetically. Rather, it can be used to select among any points of interest on a screen. The trademark of a Nomon application is a set of small clocks, one clock associated with each selectable option. Each clock appears alongside its corresponding option on the screen. For instance, Figure~\ref{fig:ng} illustrates clocks corresponding to 16 arbitrary option locations. Another example might be a drawing application where a clock appears at every ``pixel''
on the canvas and also next to each menu option. In a writing application (the Nomon Keyboard), a clock appears next to each character, word completion, or text editing function (Figure~\ref{fig:nk}).

Just as menu options and drawing tools in a point-and-click interface are accessed in the same way by the mouse, all Nomon clocks are selected in the same way by a single switch. Each Nomon clock features a moving hand and a fixed line at noon. All moving hands rotate at the same, fixed speed but, at any time, are located at a variety of angles relative to noon. The user tries to click precisely when the moving hand on her desired clock is at noon. She repeats this action until the clock is selected. Selection is signalled by the desired clock being highlighted with a darker color and the entire application flashing a lighter color; there may also be audio feedback. Between clicks (if more than one click is required to select a clock), the clock angular offsets are adjusted by a heuristic to maximize the expected information content of the user's next click.

Row-column scanning can be viewed as a special case of the Nomon selection method where clocks are arranged in a grid, moving-hand angular offsets are aligned alternately across rows and columns of clocks, and each selection is based only on the times of the last two clicks. But this synchrony does not take full advantage of the continuous, periodic representation of the clock and imposes an order on the set of options relative to their positions onscreen. Rather, by allowing more general clock hand positions, 
we can, effectively, completely reorder the set of selections after each click without demanding any extra cognitive load from the user.

Similarly, the independent movement of the clock hands frees the user to attend only to the desired target, in contrast to the need, in scanning, for the user to attend both to the desired target and the moving highlight. Further, the scanning user may forget her target as highlighting progresses. But in Nomon, once the target is located visually, the user is free (without suffering a performance penalty) to focus on selecting a single, fixed clock. Since the clock periods are usually much shorter than a full scanning rotation, there is also no significant penalty for missing a potential click time.

In Nomon, by contrast with scanning, we assume that the user will not always click perfectly at the desired time. The details of Nomon operation are described more fully in the Nomon Operation section below and outlined here. Nomon can learn a user's probability of clicking at different (typically small) offsets relative to noon. This learning is accomplished via an approximate Parzen window estimator, with contributions from more recent clicks weighted more strongly to allow adaptation to a user whose skill changes over time. We can also specify a prior probability distribution over clocks according to a predictive model of user choices. For instance, in the writing application tested below, our language model assigned prior probabilities to letters and word completions based on the British National Corpus word-frequency list~\cite{kilgarriff:1998}. These prior probabilities could also be adaptive and context dependent.

During a particular selection process, the posterior probability of any clock given the clicks thus far can be calculated from Bayes' theorem. When the probability of a single clock is sufficiently high, we declare it the winner. The probability threshold for winning is an adjustable parameter of the model; it can vary according to context or from clock to clock. A higher threshold can ensure greater safety for critical actions.

\section*{Results} 

\begin{figure}[!t]
\begin{center}
\includegraphics[scale=1]{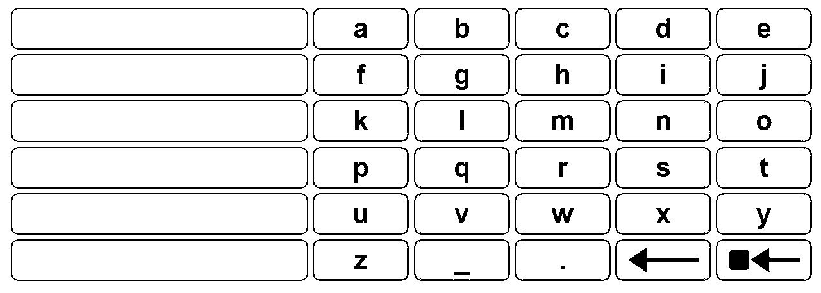}
\caption{{\bf The scanning grid from The Grid 2 used in this comparison study (screenshot).} The six long rectangles on the left hold word completions. The remaining options are fixed and include letters, an underscore for space, a period, a character-deletion function, and a word-deletion function.}\label{fig:grid}
\end{center}
\end{figure}

We developed a writing program using the Nomon method, the Nomon Keyboard (Figure~\ref{fig:nk}), and conducted a study to compare writing with Nomon to writing with a popular commercial scanning interface, The Grid 2~\cite{sensory:2008} (Figure~\ref{fig:grid}). To that end, sixteen study participants with no previous experience of either interface wrote with Nomon and The Grid 2. In each of two sessions, a participant used one of the interfaces to write short phrases appearing on screen. A session was divided into four blocks, each lasting approximately $14$ minutes. During the first three (of four) writing blocks, each participant was allowed to adjust
the rotation-period or scanning-delay parameter, as appropriate to the current interface, at the end of each written phrase. No changes were allowed during the final block. For each interface, cash prizes were won by the faster half of participants in the final block.

In total, we collected 34 hours of data from 16 novice participants and one experienced single-switch user (TB, with $>10$ hours experience in each interface).
We compared three objective measures of the novice participants’ performance between the
two interfaces: text-entry rate, error rate, and click load (clicks per character). We also examined subjective
ratings of the two interfaces given by the novice participants.

\subsection*{Text-entry Rate}

\begin{figure}[!t]
\begin{center}
\includegraphics[scale=1]{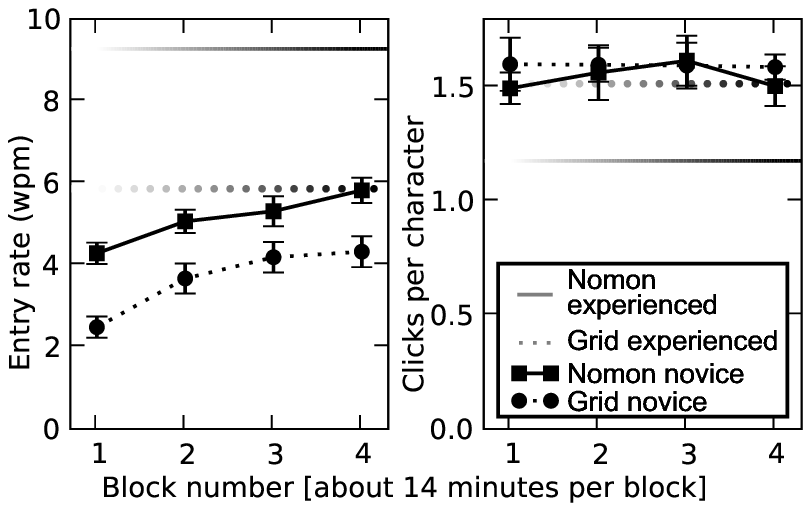}
\end{center}
\caption{{\bf Mean entry rate ({\em left}) and click load ({\em right}) across interface blocks.} Mean entry rate is measured in words per minute, and click load is measured in clicks per output character. In both panels, error bars represent
95\%
confidence intervals for the novice user means, and the average experienced user (TB) performance is illustrated by horizontal lines for comparison.}\label{fig:wpm-cpc}
\end{figure}

We calculated text-entry rate in words per minute, where a word is
defined as five consecutive characters in the output text. At the beginning of each
fourteen-minute block, the participants were asked to write two
periods ``..''
using the interface for that session. This action signalled
that they were ready to begin and initiated the display of the first target phrase. Timing
started once the two periods were written. After every phrase, participants wrote two periods to signal that they were ready for a new phrase. Timing stopped after the final
two periods following the last phrase were written. All periods except the first two in a block were counted as characters in what follows, and the time
spent writing them was counted as well.

The left panel of Figure~\ref{fig:wpm-cpc} shows the novice
participants' mean entry rates across the four blocks for each
interface. Also shown, for comparison, is the performance of the
experienced user. Participants wrote faster with Nomon than with The Grid 2
during the first block ($F_{1,15} = 129$, $p = 9.3\cdot 10^{-9}$).
The total session time was short for both interfaces, but participants' writing speed with each interface improved with practice.
Participants became faster at writing using the Nomon Keyboard during the Nomon session ($F_{3,45}=59$, $p = 1.4\cdot 10^{-15}$)
and became faster at writing using The Grid 2 during the scanning session ($F_{3,45}=122$, $p < 10^{-15}$).
In the final block we see that participants remained faster at writing with
Nomon than with The Grid 2 ($F_{1,15}=135$, $p = 6.8\cdot 10^{-9}$).
In this fourth block, participants wrote $35\%$
faster with Nomon than with the scanning interface; participants were writing at $4.3$ words per minute on average with The Grid 2 and
$5.8$ words per minute with the Nomon Keyboard. The experienced user wrote, on average, at $9.3$ words per minute with the Nomon Keyboard and $5.9$ words per minute with The Grid 2.

While the alphabetic layout was easy for novices to use, a computer simulating writing from a conversational corpus with no errors has been shown to achieve a $19\%$
faster writing speed with a frequency-ordered layout than with an alphabetic layout~\cite{venkatagiri:1999}. Even if we artificially inflate the novice writing speeds using The Grid 2 by $19\%$,
novices remain faster at writing with Nomon ($F_{1,15}=19.14$, $p = 5.4 \cdot 10^{-4}$).

\subsection*{Error Rate}

To find the error rate during a block, we begin by computing the character-level
Levenshtein distance~\cite{wagner:1974} $d_{i}$ between the $i^{\rm th}$
target phrase in the block and the text written by the participant; $d_{i}$ is also known as the edit distance.
We define the error rate for the block to be $\sum_{i} d_{i}/\sum_{i} n_{i}$, where $n_{i}$ is the number of characters in the $i^{\rm th}$ target phrase.

The average novice character-level error rate (over all blocks) for the Nomon Keyboard was $0.43\%$,
and the average novice error rate for The Grid 2 was $0.34\%$.
There was no significant difference in novice error
rate between the two interfaces ($F_{1,15}=0.71$, $p=0.41$).
The experienced user made no errors while using Nomon ($\sum_{i} d_{i} = 0$) and made one error while using The Grid 2, for a mean scanning block error rate of $0.06\%$.

We believe that the participants' output errors were mostly caused by
poor recall of the target sentence. For instance,
one participant pluralized ``head''
in ``head\_\+shoulders\_\+knees\_\+and\_\+toes'' and wrote
``reading\_\+week\_\+is\_\+almost\_\+here'' instead of
``reading\_\+week\_\+is\_\+just\_\+about\_\+here''.

\subsection*{Click Load}

The click load is the number of clicks per output-text character. Other names for this measure include ``keystrokes per character''~\cite{mackenzie:2002}
and ``gestures per character''~\cite{ward:2000}.
The click load is calculated as the number of button presses in a
block divided by the number of characters in the output. Clicking often can be tiring for any user and especially so for some users with specific motor impairments.

While the inclusion of word-completion options in a scanning grid has been shown to have no positive effect on writing speed with a scanning interface~\cite{koester:1994}, other studies confirm that word completion options yield substantial click-rate savings over the baseline (mistake-free) row-column click load of two clicks per character~\cite{koester:1996,lesher:1998:optimal}. Therefore, we included six word-completion options in the leftmost row of our scanning grid (consistent with the default layouts in The Grid 2~\cite{sensory:2008}). These were ordered from top to bottom and filled in automatically by the software.

Click loads are illustrated in the right panel of Figure~\ref{fig:wpm-cpc}.
The average novice rate (over all blocks) for the Nomon Keyboard was $1.58$ clicks per character, and the average novice rate for The Grid 2 was $1.55$ clicks per character. There was no significant difference in
novice click load between the two interfaces ($F_{1,15}=0.49$,
$p=0.49$).

While the experienced user required, on average, $1.51$ clicks per character in The Grid 2, she required only $1.18$ clicks per character using the Nomon Keyboard.
For comparison, writing with the same character set on a
normal keyboard requires at least one key press for each character
and thus at least 1 click per character (possibly more due to error correction). To compare to Morse code, we find letter, space, and period frequencies directly from our phrase set.
We assume the Morse encoding of~\cite{words:2000, itur:2004}.
In this case, an error-free Morse code click load estimate is $3.0$ clicks per character. This load is over twice as high as the click load of the experienced user on the Nomon Keyboard.

\subsection*{Subjective Ratings}

We assessed novice participants' opinions with a questionnaire
immediately after writing with an interface was completed. The questionnaires for each
interface were identical (except for the name of the interface). Participants were asked to
rate how much they agreed with a series of statements on a scale from 1
(strongly disagree) to 7 (strongly agree). These statements were largely the same as those in~\cite{kristensson:2009}.
Participants were encouraged to write any thoughts
about the interfaces in an ``Open Comments'' box.

Participants' responses to selected statements are summarized in
Table~\ref{tab:subj}. Not only did participants like using the Nomon Keyboard in aggregate, but 
every participant individually liked using Nomon at least as much as The Grid 2.
Contributing factors for why the Nomon Keyboard was preferred became
apparent in the remaining responses. Participants found it easier to
select word completions and easier to correct errors with the Nomon
Keyboard. These responses corroborate our objective findings above.

While many written comments agreed with participants' numerical ratings, unique to the open comments section was the sentiment that Nomon looks unusual at first but is worth getting to know. One participant remarked, ``Surprisingly, I found this more user-friendly.''
Another noted, ``The writing system looks intimidating when it first comes up on screen but is actually very easy to use.''

\begin{table}[!t]
\caption{{\bf Subjective ratings of the two
interfaces by novice participants.}}\label{tab:subj}
\begin{center}
\begin{tabular}{lcc}
\toprule
Statement & Nomon & The Grid 2 \\
\cmidrule{2-3}
	& mean (sd) & mean (sd) \\
\midrule
I liked writing using X. & \textbf{5.6} (1.4) & 3.9 (1.5) \\
It was easy to select word completions (the, and, cat, \ldots). & \textbf{6.1} (0.7) & 4.8 (1.3) \\
It was easy to correct errors. & \textbf{4.5} (1.8) & 3.9 (1.7) \\
\bottomrule
\end{tabular}
\end{center}
\begin{flushleft} Each response to the lefthand
statements was on a scale from 1 (strongly disagree) to 7 (strongly
agree). In the questionnaires, the interface name was substituted for X. Mean responses are shown with standard deviations in
parentheses. Boldface is used to highlight the means corresponding to a more positive user
experience.
\end{flushleft}
\end{table}

\section*{Discussion}

\begin{figure}[!t]
\begin{center}
\includegraphics[scale=1]{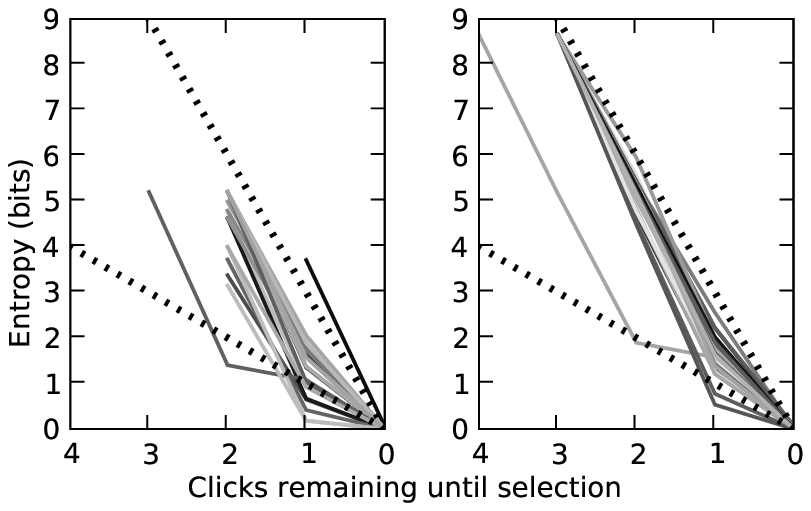}
\caption{{\bf Entropy of the estimated probability distribution over clocks for two Nomon applications.} Entropy is shown as a function of clicks remaining to selection. Each solid line represents a single selection process. Dotted lines decreasing to zero at respective rates of $1$ (lower) and $3$ (upper) bits per click are illustrated for reference. {\em Left}: 25 selections on the Nomon Keyboard: 30 clocks, non-trivial prior $p(c)$, clock period $2.0$ seconds, switch input from joystick button. {\em Right}: 25 selections on another Nomon application: 401 clocks, uniform prior $p(c)$, clock period $2.0$ seconds, switch input from space bar. Data was generated by the experienced user (TB).}\label{fig:entropies}
\end{center}
\end{figure}

Nomon benefits in this comparison from its nice scaling properties and clock-position flexibility.
Our posterior-based selection method implies that the time taken to make a selection in Nomon scales logarithmically with the number of clocks if the prior over clocks is uniform.
The entropy of the discrete uniform distribution, which happens to be the highest-entropy (finite) discrete distribution, scales logarithmically with respect to the number of points in the support. Figure~\ref{fig:entropies} shows that, generally, $2$ clicks are required by an experienced Nomon user (TB, with $>10$ hours experience) to make a selection in a 30-clock application. In a Nomon application with uniform prior and $401$ clocks, $3$ clicks are generally required for this user to make a selection. The difference in entropy between the prior for the $401$-clock application and the highest-entropy prior for the $30$-clock application is about $3.5$ bits, in agreement with $\log_{2}(401/30) = 3.7$.

Not only does the number of clicks to selection in Nomon scale well, but including additional options with small prior probabilities has little effect on clicks-to-selection for more-likely clocks.
Therefore, we could place many more word completions on screen than would be feasible for a scanning interface. We limited ourselves to three per character so as to allow fast reading of the three relevant options. Placing word completions next to letters in Nomon was feasible since clock position onscreen does not affect Nomon operation. Interspersing word completions with letters in row-column scanning would increase the number of scanning steps required to reach many options.

While a Nomon writing application allows a straightforward comparison of Nomon with existing single-switch communication methods, the Nomon selection method is not limited to writing. For example, Nomon can be used for internet browsing by placing a Nomon clock next to each link. Or Nomon can be used for drawing by placing a dense grid of, say, hundreds of clocks on a canvas. (The Nomon Draw application works in this way.) A user can draw a line by selecting points directly from the canvas. Options for colors, shape drawing, saving, and printing can likewise be accessed with clocks. A general graphical user interface can be navigated with Nomon by placing clocks at the points where a user might traditionally point and click.

It is worth pointing out that the flexibility of Nomon is not specific to our clock display choice. Other local periodic representations of the global set of options would also allow the arbitrary placement of options onscreen. For instance, the clocks could be replaced by bouncing balls at different points in their trajectories; instead of clicking at noon, the user would click when the desired ball hits the ground. It remains to be studied whether such alternative display choices might facilitate even faster or easier use of this system.

\section*{Materials and Methods}

We begin by detailing the experimental method used in the study above and follow with a description of how Nomon functions.

\subsection*{Experiment}

\subsubsection*{Participants}

We recruited sixteen participants from the university community
across a wide range of academic disciplines. All participants gave written informed consent. In accordance with the University of Cambridge ethical review procedure as defined in
the Cambridge Psychology Research Ethics Committee Handbook ({\tt http://\+www.bio.cam.ac.uk/\+sbs/\+psyres/}), the experimental design received an internal peer review within the department, where it was decided that ethics approval from the committee was not necessary.

The participants' ages ranged from
22 to 39 (mean = 26, sd = 4).
Eight were women, and eight were men.
Participants were screened for motor or cognitive difficulties; in
particular, no participant had dyslexia or RSI. None of the
participants had used a scanning or Nomon interface before. No
participant had regularly used any single-switch interface before.
Twelve of the participants had used word completion (e.g.\ on cell
phones).

In addition to the sixteen novice participants, an experienced
user of Nomon and The Grid 2 ($>10$ hours writing with each interface) was run through the same experimental
procedure for comparison.

\subsubsection*{Apparatus and Software}

All sessions were run on a Dell Latitude XT Tablet PC with a
partitioned hard drive. The 12.1 inch color screen had a physical screen
size of 261 $\times$ 163 mm. The single-switch hardware device in all
cases was the trigger button of a Logic3 Tornado USB joystick. Participants operated the trigger button with the first finger of their left or right hand.
None of the other joystick inputs was used. For both writing
interfaces, automated spoken feedback was provided as the user wrote.

\paragraph{Nomon Keyboard}
We ran the Nomon Keyboard (Figure~\ref{fig:nk}) on an
Ubuntu 8.10 operating system running the Linux kernel. The screen resolution was  1280 $\times$ 800 pixels, and the physical size of the keyboard display was 224 $\times$ 85 mm (1125 $\times$ 416 pixels). The interface was docked in the upper
part of the screen. A text box and phrase box were located below the
keyboard in the same window. The keys of the keyboard were arranged in six rows
and five columns. Each key contained a principal character,
with letters in alphabetical order (across and then
down) first, followed by four special characters: an underscore
(representing space), a period, a character-deletion function, and an
undo function. Each letter key also contained up to
three word completions. The undo function undid the previous selection if it was a character selection, word-completion selection, or deletion.

The clock rotation period $T$ could be set to
$2.0 \cdot 0.9^{j}$ seconds for $j \in \{-4,-3,\ldots,18\}$. Higher $j$ corresponded
to faster rotation. The initial setting of the period for novices was
$T=2.0$ ($j=0$). The experienced user initially chose $T=1.06$ ($j=6$).

\paragraph{The Grid 2}
We ran The Grid 2 (Figure~\ref{fig:grid}) on a Windows Vista
Service Pack 1 operating system. The screen resolution was again 1280 $\times$ 800 pixels. The physical size of The Grid 2 display, using the scanning grid we designed for this experiment, was 261 $\times$ 102 mm (1280 $\times$ 500 pixels). The interface was docked in the upper part of
the screen, and the text box and phrase box were docked immediately
below. Six word-completion boxes appeared on the left side of the
main interface. The remaining space was divided into six
rows and five columns of keys. Each key contained a
single character. First were letters arranged in alphabetical order
(across and then down), followed by an underscore, a period, a
character-deletion function, and a word-deletion function.

The Grid 2 allowed 
scanning delay values $d$ at $0.1(10-j)$ for $j \in \{\ldots,-1,0,1,\ldots,9\}$. Higher $j$ corresponded to faster scanning. The initial
setting of the delay for novices was $d = 1.0$ ($j=0$). The
experienced user initially chose $d = 0.5$ ($j=5$).

\subsubsection*{Procedure}

The experiment consisted of two identical sessions, one for each
interface. The starting interface was balanced across participants,
and sessions were spaced at least four hours apart.

Each session proceeded according to the same schedule. The first ten
minutes were introductory. First, the supervisor either explained or reviewed the experimental procedure according to the session number. Then the participant was shown how
to use one of the interfaces. The demonstration included basic
writing, word completion, and error correction.

The next hour was divided into four 14-minute blocks, separated by short breaks. During the blocks, participants were
asked to write phrases drawn from a modified version of the phrase
set provided by~\cite{mackenzie:2003}, with British spellings and
words substituted for their American counterparts. For each
participant, a different random ordering of the initial phrase set
was generated. Phrases appeared one at a time in the phrase box at
the bottom of the screen. Once a participant finished a
phrase, writing the period character twice would cause a new target phrase
to appear and the text box to empty. Participants were instructed
that no changes relevant to a particular phrase could be made after
the two periods were written.

During the first three (of four) writing blocks, each participant was allowed
to adjust the rotation-period or scanning-delay parameter at the end
of each written phrase. In particular, immediately after writing
two periods and receiving the new target phrase, the participant
could increment or decrement $j$ (defined above) by one. The experienced user incremented to $j=7$ ($T=0.96$) after two blocks using the Nomon Keyboard and incremented to $j=6$ ($d=0.4$) after two blocks using The Grid 2. No other changes were made by this user.

Novice participants were paid \textsterling 10 for each of the two sessions; the experienced participant was not paid. Novice participants were informed at the beginning of the study that they
could receive a \textsterling 5 bonus for achieving a writing
speed among the top half of novice participants for each interface. They
were further informed that, for the purposes of the bonus, writing
speed would be measured only during the final writing block. They
were told that they would not be allowed to change the rotation-period
or scanning-delay parameter during this block and thus would
have to calibrate it as they saw fit during the previous blocks.
Information about their own writing speeds across full blocks and
also phrase-by-phrase was made available to participants during
the break after each block.

We performed seven
significance tests with a family-level significance of $0.05$. Observing the Bonferroni correction, we performed each individual test at a significance level of $\alpha = 0.007$. Wherever $F$ values are quoted, an analysis of variance (ANOVA) test for repeated measures was performed.

\subsection*{Nomon Operation}

We here describe the prior over clocks, click likelihood (given a clock), and the resulting posterior over clocks in turn. While we focus on a prior for a specific application (the Nomon Keyboard), the likelihood and posterior discussions are germane to a general Nomon application.

\subsubsection*{Prior}

In the absence of information about clock probabilities, we use a uniform prior $p(c)$ over clocks $c: 1 \le c \le C$. We can choose a more informative prior for our Nomon writing application, the Nomon Keyboard (Figure~\ref{fig:nk}).
This interface features four special characters (underscore representing space; period; Delete; and Undo), 26 letters, and up to three word completions per letter. We assign fixed prior probabilities to the special characters and assign the remaining priors according to Laplace smoothing out of the leftover probability mass $p_{\rm alpha}$. Let $l_{1}\cdots l_{N}$ ($N \ge 0$) be the context (all letters from the end of the current output text) before the user begins to make another selection. Let $\mathcal{W}_{\rm on}$ be the set of word completions appearing on screen, and set $C_{\rm on} = |\mathcal{W}_{\rm on}| + 26$.
To form our corpus, we begin with the British National Corpus word list~\cite{kilgarriff:1998}, then we remove single-letter words besides ``I'' and ``a''
and keep only words appearing with some small minimum frequency ($>5$ appearances in the corpus).

When an appropriate word completion is offered, the user may nevertheless choose the next single letter; the following model assumes that the user is equally likely to choose either of these options.
If $f_{w}$ is the number of occurrences of word $w$ in our corpus, we define a context frequency $f(l_{1}\cdots l_{N}) = \sum_{w} f_{w} \mathbbm{1}\{l_{1}\cdots l_{n} \textrm{ prefixes } w\}$ and a screen word-completion summed frequency $f(\mathcal{W}_{\rm on}) = \sum_{w' \in \mathcal{W}_{\rm on}} f(w')$. If $c(l')$ is the clock corresponding to letter $l'$ and $c(w)$ the clock corresponding to word $w$,
\begin{eqnarray}
p(c(l')) &=& p_{\rm alpha} \times \frac{f(l_{1}\cdots l_{N} l')+1}{f(l_{1}\cdots l_{N})+f(\mathcal{W}_{\rm on}) + C_{\rm on}} \\
p(c(w)) &=& p_{\rm alpha} \times \frac{f(w)+1}{f(l_{1}\cdots l_{N})+f(\mathcal{W}_{\rm on}) + C_{\rm on}}
\end{eqnarray}
To model an ideal user, we would subtract the count of words onscreen prefixed by $l_{1}\cdots l_{N} l'$ from the numerator of $p(c(l'))$, and both denominators would equal $f(l_{1}\cdots l_{N}) + C_{\rm on}$.
Finally, while the number of letters is fixed at 26, $C_{\rm on}$ is variable since, for any letter, we include only those word completions among the three most probable above a certain threshold. It was judged that requiring $f_{w}/f(l_{1}\cdots l_{N}) > 0.001$ yielded a reasonable balance between displaying common words and not cluttering the screen.

\subsubsection*{Click Distribution}

Any particular clock $c$ defines a desired click time at noon. We wish to estimate a user's click time distribution relative to noon $p(t|c)$, where we distinguish $t$ only up to the clock period $T$ and set $t|c$ to zero at noon. To that end, we begin with a broad, and slightly offset, initial setting of our estimate for $p(t|c)$: $\hat{g}_{0}(t) = \mathcal{N}(t;0.05T,(0.14T)^{2})$. The $T$-dependence ensures the estimate will be nontrivial at any user-chosen period. We update the $\hat{g}_{0}$
distribution with a (modified) Parzen window estimator---with width given below---and a damping factor
$\lambda$ that allows learning to continue over time.
After any selection is made, we modify the distribution estimate with
the data from the $n_{\rm delay}^{\rm th}$ selection before the latest one (here $n_{\rm delay} = 2$).
This delay allows the user to choose \texttt{Undo} after a
selection, in which case we do not use the clicks toward that
selection for learning. Once a selection occurred $n_{\rm delay}$ rounds
in the past, we assume that it was correctly chosen. With the clock
choice $c$ known for the $s^{\rm th}$ selection, we are able to calculate
click times around noon $t_{s,r}$ for each click that was made
toward this selection. We treat these as data from the distribution
$g$ we are estimating. To calculate our estimate $\hat{g}_{s}$ for $g$ after
the $s^{\rm th}$ selection, we make use of the unnormalized distributions $\tilde{g}_{s}$.
\begin{equation}
	\tilde{g}_{s}(t) = \lambda \tilde{g}_{s-1}(t) + \sum_{r=1}^{R_{s}} \mathcal{N}\left(t;t_{s,r},\hat{\sigma}^{2}_{\rm{NS},s}\right) \textrm{ with } \tilde{g}_{0}(t) = n_{\lambda} \hat{g}_{0}(t)
\end{equation}
The update equation specifies that, after each selection, $\tilde{g}$ is damped by the factor
$\lambda$. The next term is a sum over clicks $r$ leading to the $s^{\rm th}$ selection. Within
the summation is a normal density centered at the click time
$t_{s,r}$, as in Parzen window estimation. The width for this Parzen-window term is given by $\hat{\sigma}_{\textrm{NS},s}$, which
is derived from the normal scale rule estimate~\cite{wand:1995,silverman:1986} for the Parzen window. That is,
\begin{equation}
	\hat{\sigma}_{\textrm{NS},s} = 1.06 n_{\lambda}^{-0.2} \hat{\sigma}_{s},
\end{equation}
where $\hat{\sigma}^{2}_{s}$ is the standard
(Gaussian maximum likelihood) variance estimator obtained from the last $n_{\lambda}$ clicks
before the $s^{\rm th}$ round. The factor $n_{\lambda} = (1-\lambda)^{-1}$ in the initial $\tilde{g}_{0}$ definition is an effective number of
samples derived from the damping factor. Using this factor and the unnormalized update, we ensure that the initial estimate $\hat{g}_{0}$ dominates $\hat{g}_{s}$ even after the first few selections. Without the $n_{\lambda}$ factor, the Parzen window term for the first click, $\mathcal{N}\left(t;t_{1,1},\hat{\sigma}^{2}_{\rm{NS},1}\right)$, would have nearly equal weight with the initial estimate.

This estimate for $p(t|c)$ allows us to save the estimated distribution and update it quickly
and easily during operation of the application. As a result, users
can start the Nomon application immediately, without a waiting or
calibration period, but they can also enjoy an experience
tailored to their abilities. For instance, a user need not click at noon (or any offset) exactly. Their personal
offset, reflecting reaction time, is learned by this method rather
than hard-coded and, as long as it is not too close to 6 o'clock, will make no
difference to program operation. The precision around this
personal offset determines the number of clicks necessary to
make a selection.

\subsubsection*{Posterior}

With a prior and likelihood, we may calculate the posterior probability
of each clock $c$ given the $R$ clicks thus far using Bayes' theorem: $p_{c,R}
= p(c|t_{1:R}) \propto p(c) \prod_{r=1}^{R} p(t_{r}|c)$. In practice, we store the unnormalized log probabilities for each $p_{c,R}$. Checking that the highest clock probability $p_{(C),R}$ exceeds some threshold would require exponentiating every stored value and summing over the results. Noting that $p_{(C),R} > 1 - p_{\rm error}$ is equivalent to $p_{(C),R} > \alpha \sum_{c\ne (C)} p_{c,R}$ for some $\alpha$, we instead declare a winner when $p_{(C),R} > \alpha p_{(C-1),R}$. The choice of $\alpha = 99$ represents a desired upper bound on error fraction, {\em per selection}, of $0.01$.
In a sample of 1,714 consecutive selections made by an experienced Nomon user (TB) on the Nomon Keyboard under this setting, the average value of $p_{(C),R}/p_{(C-1),R}$ over all selections after the deciding click was $0.001$, and the average value of $p_{(C),R}/\sum_{c\ne (C)} p_{c,R}$ was $0.002$, suggesting our heuristic stopping criterion is a reasonable approximation to the desired one. In the 1,714 selections, 3 (non-consecutive) selections were \texttt{Undo}, indicating mistakes and giving an empirical error rate of about $0.002$, in line with the calculated rate.

\section*{Acknowledgments}
This research was supported by donations from the Nine Tuna Foundation and Nokia.
We thank Sensory Software for generously providing the 60-day free trial of The Grid 2 used in our performance comparison. We are grateful to Per Ola Kristensson, Geoffrey Hinton, Keith Vertanen, Philipp Hennig, Carl Scheffler, and Philip Sterne for helpful discussions.
TB's research is supported by a Marshall Scholarship.


\end{document}